# Career Path Suggestion using String Matching and Decision Trees

Akshay Nagpal
Dept. of Computer Science
ITM University
Sector 23-A, Gurgaon, India

Supriya P. Panda
Dept. of Computer Science
ITM University
Sector 23-A, Gurgaon, India

## ABSTRACT
High school and college graduates seemingly are often battling for the courses they should major in order to achieve their target career. In this paper, we worked on suggesting a career path to a graduate to reach his/her dream career given the current educational status. Firstly, we collected the career data of professionals and academicians from various career fields and compiled the data set by using the necessary information from the data. Further, this was used as the basis to suggest the most appropriate career path for the person given his/her current educational status. Decision trees and string matching algorithms were employed to suggest the appropriate career path for a person. Finally, an analysis of the result has been done directing to further improvements in the model.

## General Terms
Decision Trees [4], Supervised Machine Learning, String Matching, Career Suggestion

## 1. INTRODUCTION
It is always a daunting task to decide a career goal for High school or College graduates with selected courses, which will take them to their ambition in the end. At this juncture, they start glancing at profiles of allied people and start comparing with their own. But this comparison is limited to only their acquaintances. This model was developed to solve this very problem faced by the students. This model compares the student's career goal to thousands of other career paths collected through online surveys and LinkedIn. This facilitates the student to know the steps he/she needs to take further to reach his/her career goal. At the input level, the student needs to feed current education and his/her career goal. At the output level, the various career paths he/she can opt to achieve his goal would be generated.

## 2. DATA
### 2.1 Data Collection
The career data for the model was collected online through LinkedIn and surveys. Linked data from various profiles was extracted using LinkedIn API [5]. Google Forms were used to create forms to be filled by professionals of various industries.

### 2.2 Data Organization
Data was organized into a Comma Separated Value (CSV) file so that it can be accessed easily through Python.

The data had the following attributes:

i. Bachelors Stream
ii. Bachelors University
iii. Bachelors Duration
iv. Masters Stream
v. Masters University
vi. Masters Duration
vii. Doctoral Stream
viii. Doctoral University
ix. Doctoral Duration
x. Work Position
xi. Work Organization

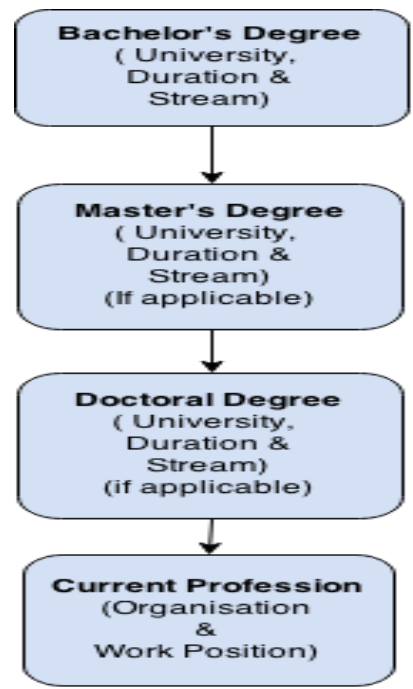

**Fig.1 Order of Career Path of a Person**

Person's career path is viewed in the order as depicted in Fig.1.

## 3. ASSUMPTIONS
To simplify the model, we assume that:

1. The user's current job depends only on his most recent education.
2. The user is not sure of his/her career path if he/she is below the Master's level, i.e., pursuing High School or Bachelor's.

Apparently, Masters itself contains specializations which the user must have opted after deciding his/her career path, so there are less chances of him/her being confused about the career path.

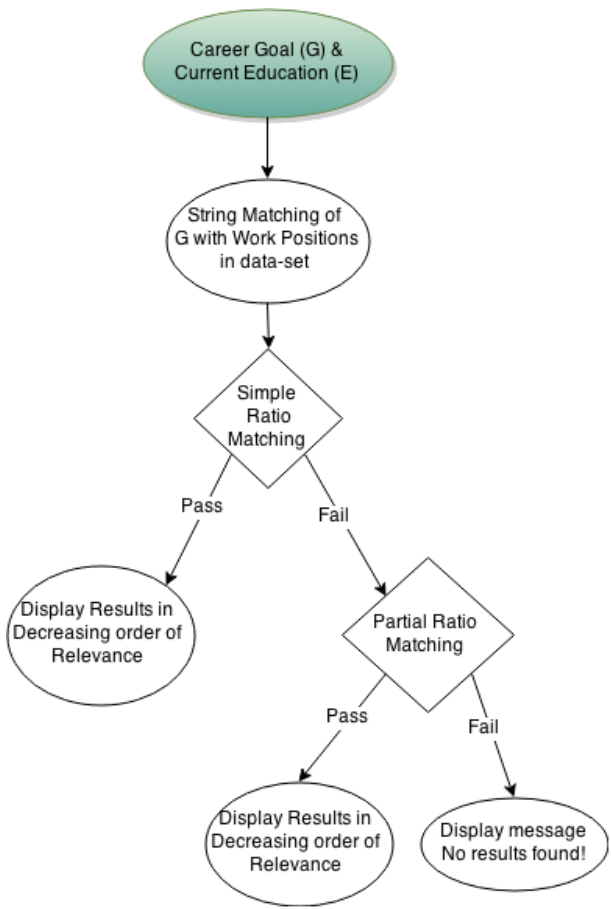

**Fig.2 Suggestion Mechanism**

## 3.1 String Matching
After the user inputs the details, string matching of Career Goal (G) is done as depicted in Fig.2 with the values in the "Work Position" (WP) attribute of the data set. We have used fuzzy string matching using the Fuzzywuzzy library in Python.[1]

We have used two of the **string matching methods** [3] provided by the library as follows:

   a. Simple Ratio : ratio(s1,s2)

   The two strings s1 and s2 are of identical length that has to be compared.

   The result is the similarity between the two strings given in percentage.

   If T = total number of elements in both sequences, and

   M = number of matches,

   Similarity Percentage (SP) is calculated using the formula: SP=200*(M / T).

   $$SP = \begin{cases} 100 & \text{if the sequences are identical} \\ 0 & \text{if they have nothing in common} \end{cases}$$

   b. Partial Ratio : partial_ratio(s1,s2)

   If the two strings s1, s2 are of different lengths, e.g. 'Developer' and 'Software Developer', partial_ratio method (PM) is used.

   If the length of shorter string = m, and the length of longer string = n, then the method finds the most similar substring of length m from both the strings and applies simple ratio (SR) method on them.

## 3.2 Career Path Suggestion Algorithm
A Suggestion Algorithm was developed which is described below.

**Career Path Suggestion (G, E)**

*Input:* the career goal G of the user and **E** the current education.

*Output:* the list of various career paths **path_list** [ ]  suggested to the user.

{

1. **G** is a string that can have any value as there are various careers that the user can input.
   **E** can have the value "High School" or "Bachelor's" depending on the student's current education.
2. Let path_**list** [ ] be the list of various career paths suggested to the user by the result of the model; initially, **path_list**[ ] = empty.
3. For each row in dataset, simple ratio matching of **G** and **WP** is done. If similarity percentage (SP) between **G** and **WP** is greater than 60, then simple ratio match is successful.
   (a) If **E** = "High School", add Bachelor's information, Master's information and Doctoral information to the **path_list[ ]**.
   (b) Else if **E** = "Bachelor's", add Master's information and Doctoral information to the **path_list**[].
   (c) Else, simple ratio (SR) matching is failed.
4. If simple ratio matching fails, then for each row in dataset, execute partial ratio matching between **G** and **WP**. If similarity percentage (SP) after partial ratio matching is greater than 80, partial ratio matching is successful. Add row to the **path_list[]**.
5. Sort resulting list in decreasing order of similarity.
6. Display **path_list**[] of resulting career path suggestions.

}

## 4. RESULTS
When given an input, our model gave logical and real world career suggestions as shown in Fig.3 below.

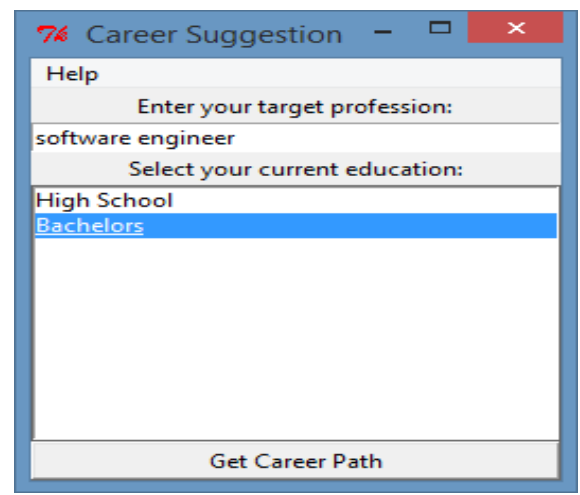

**Fig.3 Graphical User Interface (User Input)**

For example, when the user entered "Software Engineer" in the text field and selected "Bachelor's" in the Current Education option, the model gave the following result as in Fig.4 below:

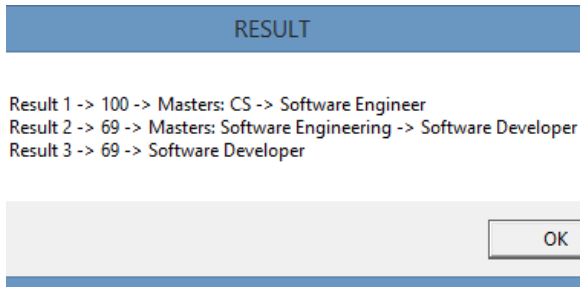

**Fig.4 Graphical User Interface (Display of Result)**

The above result is corresponding to the human common sense and any human career counselor would have given the above three alternatives to any person asking a similar doubt.

Table 1 shows some inputs given to the model and the corresponding career paths given as an output.

**Table 1: Results of some random inputs given to the model**

| S.No. | Input (Target, Current Education) | Output (Career Path) |
|---|---|---|
| 1 | Software Engineer, Bachelor's | Masters , Computer Science |
| | | Masters , Software Engineering |
| 2 | Fashion Designer, High School | Bachelors, Fashion Designing |
| | | Bachelors, Fashion Merchandising |
| 3 | English Professor, High School | Bachelors, English > Masters, English Literature |
| | | Bachelors, English > Masters, English Literature > Doctoral, English Language & Literature |
| 4 | Data Scientist, Bachelor's | Masters, Computer Science > Doctoral, Statistics |
| | | Masters, Information Technology > Doctoral, Data Science |
| | | Masters, Data Science |

## 6. CONCLUSION

Our model offers real world results for basic and intermediate queries of the students. However, it can be made more efficient to handle other advanced and complex searches.

Firstly, the ranking of the universities in the career paths can be used to rank the different paths; the career paths with the better ranked universities will come at the top in the search results which will make the search more convenient for the users.

Secondly, the internships done during the Bachelors or Masters can also be included to better analyze the relation between the kind of job role at the internship and the final job after graduation or post-graduation. This will give the students an idea about the kind of internships they need to take up in order to reach their career goal.